\documentclass[conference]{IEEEtran}

\usepackage[utf8]{inputenc}
\usepackage[T1]{fontenc}
\PassOptionsToPackage{hyphens}{url}\usepackage{hyperref}
\usepackage{microtype}
\usepackage{textcomp}
\usepackage{fancyvrb}
\usepackage{amsfonts}
\usepackage{amsmath}

\usepackage{acronym}

\usepackage{xcolor}
\usepackage{subcaption}
\usepackage{graphicx}
\usepackage{stfloats}
\usepackage{tikz}
\usetikzlibrary{decorations.pathreplacing}

\usepackage{array}
\usepackage{booktabs}
\usepackage{tabularx}

\usepackage[noend, noline, linesnumbered, ruled]{algorithm2e}
\SetCommentSty{small}
\SetInd{0.5em}{0.5em}

\let\oldnl\nl \newcommand{\nonl}{\renewcommand{\nl}{\let\nl\oldnl}}

\usepackage{listings}
\lstdefinelanguage{scala}{
  morekeywords={abstract,case,catch,class,def,do,else,extends,false,final,finally,for,forSome,if,implicit,import,lazy,match,new,null,object,override,package,private,protected,return,sealed,super,this,throw,trait,true,try,type,val,var,while,with,yield},
  otherkeywords={=>,<-,<\%,<:,>:,\#},
  sensitive=true,
  morecomment=[l]{//},
  morecomment=[n]{/*}{*/},
  morestring=[b]",
  morestring=[b]',
  morestring=[b]"""
}[keywords,comments,strings]

\DeclareMathOperator{\avg}{avg}

\newcommand{\sys}{MDB}
\newcommand{\sysg}{MDB\textsubscript{+}}
\newcommand{\sysgona}{\sysg{}$+$GA}
\newcommand{\sysgonb}{\sysg{}$+$GB}
\newcommand{\sysgoff}{\sysg{}$-$G}
\newcommand{\core}{\sys{} Core}
\newcommand{\coreg}{\sysg{} Core}
\newcommand{\tss}{\mathbb{TS}}
\newcommand{\algo}{GOLEMM}

\acrodef{dbms}[DBMS]{Database Management System}
\acrodef{rdbms}[RDBMS]{Relational Database Management System}
\acrodef{tsms}[TSMS]{Time Series Management System}
\acrodef{olap}[OLAP]{Online Analytical Processing}
\acrodef{oltp}[OLTP]{Online Transaction Processing}
\acrodef{lca}[LCA]{Lowest Common Ancestor}

\acrodef{pmc}[PMC-Mean]{the constant PMC-Mean model}
\acrodef{swing}[Swing]{the linear Swing model}
\acrodef{fb}[Gorilla]{Gorilla's lossless floating-point compression algorithm}

\acrodef{udf}[UDF]{User-Defined Function}
\acrodef{udaf}[UDAF]{User-Defined Aggregate Function}

\acrodef{mmc}[MMC]{Multi-Model Compression}
\acrodef{mgc}[MGC]{Model-based Group Compression}
\acrodef{mmgc}[MMGC]{Multi-Model Group Compression}
 \usepackage{cite}
\usepackage{amsmath,amssymb,amsfonts}
\usepackage{algorithmic}
\usepackage{graphicx}
\usepackage{textcomp}
\usepackage{xcolor}
\def\BibTeX{{\rm B\kern-.05em{\sc i\kern-.025em b}\kern-.08em
T\kern-.1667em\lower.7ex\hbox{E}\kern-.125emX}}
\begin{document}

\nocite{db:2018:modelardb}
\nocite{rw:2014:cloud}
\nocite{rw:2013:wsn}
\nocite{rw:2014:cloud}
\nocite{rw:2018:compression:survey}
\nocite{rw:2018:compression:survey}
\nocite{rw:2013:model:survey}

\pagenumbering{gobble}
\markboth{\begin{minipage}[b]{\textwidth}\normalsize{}Preprint of: S. K. Jensen, T. B. Pedersen, and C. Thomsen, ``Scalable Model-Based Management of Correlated Dimensional Time Series in ModelarDB\textsubscript{+},'' \textit{IEEE 37th ICDE}, 2021, pp. 1380-1391, doi: \href{https://www.doi.org/10.1109/ICDE51399.2021.00123}{10.1109/ICDE51399.2021.00123}, Copyright IEEE.\end{minipage}}{}

\title{Scalable Model-Based Management of Correlated Dimensional Time Series in ModelarDB\textsubscript{+}}

\author{\IEEEauthorblockN{Søren Kejser Jensen}
    \IEEEauthorblockA{\textit{Aalborg University, Denmark}\\
    skj@cs.aau.dk}
    \and
    \IEEEauthorblockN{Torben Bach Pedersen}
    \IEEEauthorblockA{\textit{Aalborg University, Denmark}\\
    tbp@cs.aau.dk}
    \and
    \IEEEauthorblockN{Christian Thomsen}
    \IEEEauthorblockA{\textit{Aalborg University, Denmark}\\
    chr@cs.aau.dk}
}

\maketitle

\begin{abstract}
    To monitor critical infrastructure, high quality sensors sampled at a high frequency are increasingly used. However, as they produce huge amounts of data, only simple aggregates are stored. This removes outliers and fluctuations that could indicate problems. As a remedy, we present a model-based approach for managing time series with dimensions that exploits correlation in and among time series. Specifically, we propose compressing groups of correlated time series using an extensible set of model types within a user-defined error bound (possibly zero). We name this new category of model-based compression methods for time series \emph{\acf{mmgc}}. We present the first \ac{mmgc} method \emph{\algo{}} and extend model types to compress time series groups. We propose primitives for users to effectively define groups for differently sized data sets, and based on these, an automated grouping method using only the time series dimensions. We propose algorithms for executing simple and multi-dimensional aggregate queries on models. Last, we implement our methods in the \acf{tsms} ModelarDB (\emph{ModelarDB\textsubscript{+}}). Our evaluation shows that compared to widely used formats, ModelarDB\textsubscript{+} provides up to 13.7 times faster ingestion due to high compression, 113 times better compression due to the adaptivity of \algo{}, 630 times faster aggregates by using models, and close to linear scalability. It is also extensible and supports online query processing.
\end{abstract}

\section{Introduction}\label{sec:intro}
To maintain a high output from energy producing entities, such as wind turbines, they are monitored by regularly sampling high quality sensors with wired power and connectivity. Thus, invalid, missing and out-of-order readings are rare, and all but missing values can be corrected using established methods. The data is used by park owners and turbine manufactures for management and warranty purposes. From discussions with both, we learned that storing the raw sensor data is either infeasible or prohibitively expensive. Instead, only simple aggregates are stored, e.g., 1--10 minute averages, which remove informative outliers and fluctuations.
As a \emph{running example}, consider a company with 16 wind parks spread over 3 countries. A park contains 12 turbines on average and each turbine is monitored by 98 sensors with a 100 ms sampling interval ($SI$). While a data point is only 96 bits (timestamp and value), \emph{44 TiB are stored per month}. Metadata, e.g., location, is also stored for each time series to support analysis along multiple dimensions.
Using a short $SI$ only during critical events could reduce the storage required. But what constitutes a critical event is often not known in advance so a short $SI$ is always needed.

\begin{figure}
  \centering
  \includegraphics[width=\columnwidth]{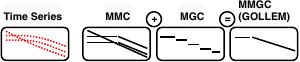}
  \caption{Our novel \acf{mmgc} method GOLLEMM gives the benefits of \acf{mmc} and \acf{mgc}}\label{fig:motivation}
\end{figure}

As a remedy, sensor data can be efficiently stored within a \emph{known error bound} (possibly zero) using \emph{models}~\cite{rw:2013:model:survey, rw:2013:model:evaluation}. Many state-of-the-art model-based compression methods use \acf{mmc} or \acf{mgc}. \ac{mmc} methods compress each time series using multiple model types to adapt as each time series' structure changes over time~\cite{rw:multi:2011:towards, rw:multi:2015:indexable, rw:multi:2015:grid, db:2018:modelardb, rw:multi:2020:mi}. \ac{mgc} methods compress correlated time series as one stream of models to not store similar values from multiple time series~\cite{rw:2009:gamps, rw:2018:mtsc, rw:2019:corad}. However, no \ac{mmc} methods exploit that time series are correlated, and each \ac{mgc} method only uses one model type. We propose combining \ac{mmc} and \ac{mgc} to reduce the models (and bytes) required to store a group of time series as shown in Figure~\ref{fig:motivation}. We name this new category of model-based compression methods \emph{\acfi{mmgc}}. To design an \ac{mmgc} method, multiple questions must be answered. (i) How can a group of time series be compressed more using multiple model types? (ii) How should time series be grouped when using multiple model types? (iii) How can existing model types be used? (iv) How can queries be executed on models? To answer these we propose: The first \ac{mmgc} method \emph{\algo{}}, an acronym for \emph{Group Online Lossy and lossless Extensible Multi-Model compression}. Primitives so users effectively can group time series either manually or automatically. Algorithms for executing aggregates on models. To show our methods are practical we add them to the \ac{tsms} \emph{ModelarDB} (MDB)~\cite{db:2018:modelardb}. \emph{ModelarDB\textsubscript{+} (\sysg{})} is the new version. \sysg{} is distributed, scales, supports online analytics, and is extensible. Compared to popular formats \sysg{} has up to 13.7x faster ingestion, 113x better compression, and 630x faster aggregation. In summary, we make these contributions in the area of big data systems:

(i) The \acf{mmgc} category of model-based compression methods for time series.

(ii) \algo{}, a \aclu{mmgc} method and model types extended to compress time series groups.

(iii) Primitives for users to effectively group time series, and a method that automatically groups them using their dimensions.

(iv) Algorithms for executing simple and multi-dimensional aggregate queries on models representing time series groups.

(v) \sysg{}'s source-code \url{people.cs.aau.dk/~skj/MDB+.tar.gz}.

(vi) An evaluation of \sysg{} on real-life and derived data.

The structure of the paper is as follows. Definitions are provided in Section~\ref{sec:prem}. Section~\ref{sec:arch} describes how \sysg{} supports \algo{}. Section~\ref{sec:partitioning} documents our grouping primitives and our automated grouping method, while Section~\ref{sec:compression} describes our extended model types. In Section~\ref{sec:query} our query processing algorithms are described. An evaluation of \sysg{} is given in Section~\ref{sec:evaluation}. Related work is presented in Section~\ref{sec:rw}. Last, Section~\ref{sec:conclusion} provides our conclusion and future work.
 \section{Preliminaries}\label{sec:prem}
As we build upon~\cite{db:2018:modelardb}, updated definitions from it are used with Model Type, Dimension, and Time Series Group added.

\textbf{Time Series:}\label{def:ts}
A \emph{time series} TS is a sequence of \emph{data points}, in the form of timestamp and value pairs, ordered by time in increasing order $TS = \langle (t_1, v_1), (t_2, v_2), \ldots \rangle$. For each pair $(t_i, v_i)$, $1 \leq i$, the timestamp $t_i$ represents when the value $v_i \in \mathbb{R}$ was recorded. A time series $TS = \langle (t_1, v_1), \ldots, (t_n, v_n) \rangle$ with a fixed number of $n$ data points is a \emph{bounded time series}.

\textbf{Regular Time Series:}
A time series $TS = \langle (t_1, v_1), (t_2, v_2), \ldots \rangle$ is considered \emph{regular} if the time elapsed between each data point is always the same, i.e., $t_{i+1} - t_i = t_{i+2} - t_{i+1}$ for $1 \leq i$ and \emph{irregular} otherwise.

\textbf{Sampling Interval:}
The \emph{sampling interval} SI of a regular time series $TS = \langle (t_1, v_1), (t_2, v_2), \ldots \rangle$ is the time elapsed between each pair of consecutive data points in the time series: $SI = t_{i + 1} - t_i$ for $1 \leq i$.

Consider, e.g., a time series with wind speed in m/s $TS = \langle (100, 9.43), (200, 9.09), (300, 8.96), (400, 8.62),(500, 8.50),$ $\ldots \rangle$. The time\-stamps are the ms since recording began. The interval $100 \leq t_i \leq 500$ of $TS$ is an example of a bounded time series. Both time series are regular with a $SI$ of $100$ ms.

\textbf{Model:}\label{def:model}
A \emph{model} of a time series $TS = \langle (t_1, v_1), \ldots \rangle$ is a function $m$. For each $t_i$, $1 \leq i$, $m$ is a real-valued mapping from $t_i$ to an estimate of the value $v_i$ for the corresponding data point in $TS$.

\textbf{Model Type:}\label{def:modeltype}
A \emph{model type} is pair of functions $M_T = (m_t, e_t)$. $m_t(TS, \epsilon)$ is a partial function, which when defined for a bounded time series $TS$ and a non-negative real number $\epsilon$ returns a model $m$ of $TS$ such that $e_t(TS, m) \leq \epsilon$. $e_t$ is a mapping from $TS$ and $m$ to a non-negative real number representing the error of the values estimated by $m$. We call $\epsilon$ the \emph{error bound}.

Intuitively, linear regression is a model type while a linear function is a model. A model type defines how to represent a time series as a model $m$, how the error of $m$ is calculated ($e_t$), and how to interpret the error bound $\epsilon$. We say a model is \emph{fitted} to a time series when its parameters are computed. The result of fitting a linear function to $TS$ within $\epsilon = 0.25$ according to the uniform norm could, e.g., be $m = -0.003t_i + 9.8$, $100 \leq t_i \leq 500$ as its error is $|9.09 - (-0.003 \times 200 + 9.8)| = 0.11$.

\textbf{Gap:}\label{def:gap}
A \emph{gap} between a regular bounded time series $TS_1 = \langle (t_{1}, v_{1}), \ldots, (t_{s}, v_{s}) \rangle$ and a regular time series $TS_2 = \langle (t_{e}, v_{e}), (t_{e+1}, v_{e+1}), \ldots \rangle$ with the same sampling interval $SI$ and recorded from the same source, is a pair of timestamps $G = (t_s, t_e)$ with $t_e = t_s + m \times SI$, $m \in \mathbb N_{\geq 2}$, and where no data points exist between $t_s$ and $t_e$.

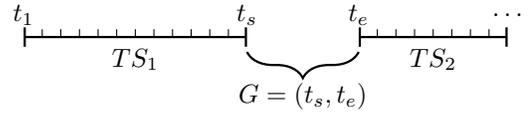
\begin{figure}
  \centering
  \scalebox{0.99}{\begin{tikzpicture}
      \draw node[midway,yshift=0.3cm] {$t_1$};
      \draw node[midway,xshift=3cm, yshift=0.3cm] {$t_s$};
      \draw node[midway,xshift=4.5cm, yshift=0.3cm] {$t_e$};
      \draw node[midway,xshift=6.5cm, yshift=0.3cm] {$\ldots$};

      \draw[|-|,line width=0.30mm,color=black] (0,0) -- (3,0);
      \foreach \x in {0.25,0.5,0.75,1,1.25,1.5,1.75,2.0,2.25,2.5,2.75} \draw[shift={(\x,0)},color=black] (0pt, 0pt) -- (0pt, 3pt);

      \draw [decorate,line width=0.30mm,decoration={brace,amplitude=10pt,mirror}](3,-0.2) -- (4.5,-0.2) node[midway,yshift=-0.6cm] {$G = (t_s, t_e)$};

      \draw[|-|,line width=0.30mm,color=black] (4.5,0) -- (6.5,0);
      \foreach \x in {4.75,5,5.25,5.5,5.75,6.0,6.25} \draw[shift={(\x,0)},color=black] (0pt, 0pt) -- (0pt, 3pt);

      \draw node[midway,xshift=1.5cm, yshift=-0.3cm] {$TS_1$};
      \draw node[midway,xshift=5.5cm, yshift=-0.3cm] {$TS_2$};
    \end{tikzpicture}}
  \caption{A gap $G$ between timestamps $t_s$ and $t_e$~\protect\cite{db:2018:modelardb}}\label{fig:gap}
\end{figure}

\textbf{Regular Time Series with Gaps:}\label{def:rtswg}
A \emph{regular time series with gaps} is a regular time series, $TS = \langle (t_1, v_1), (t_2, v_2), \ldots \rangle$ where $v_i \in \mathbb{R} \cup \{\bot\}$ for $1 \leq i$. All sub-sequences in $TS$ of the form $\langle (t_s, v_s), (t_{s + 1}, \bot), \ldots,  (t_{e - 1}, \bot), (t_e, v_e) \rangle$ where $v_s, v_e \in \mathbb{R}$, are denoted as \emph{gaps} $G=(t_s, t_e)$.

A gap is shown in Figure~\ref{fig:gap}. Time series from one source separated by gaps we refer to as a time series with gaps. For example, $TS_g = \langle (100, 9.43), (200, 9.09), (300, 8.96), (400,$ $8.62), (500, 8.50), (1100, 7.08), \ldots \rangle$ has a gap $G = (500,$ $1100)$, is irregular, and has an undefined $SI$. However, as a regular time series with gaps $TS_{rg} = \langle (100, 9.43), (200, 9.09),$ $(300, 8.96), (400, 8.62), (500, 8.50), (600, \bot), (700, \bot), (800,$ $\bot), (900, \bot), (1000, \bot), (1100, 7.08), \ldots \rangle$ its $SI = 100$ ms.

\textbf{Dimension:}
A \emph{dimension} with members $M$ is a 3-tuple $D = (\mathit{member} : \tss{} \to M,  \mathit{level} : M \to \{0, 1, \ldots, n\}, \mathit{parent} : M \to M)$ where \emph{(i)} $M$ is hierarchically organized descriptions of the time series in the set of time series $\tss{}$ with the special value $\top \in M$ as the top element of the hierarchy; \emph{(ii)} $\mathit{level}$ is surjective; \emph{(iii)} For $TS \in \tss{}$, $\mathit{level}(\mathit{member}(TS)) = n$ and $\nexists m \in M$ where $\mathit{level}(m) > n$; \emph{(iv)} For $TS \in \tss{}$, $m \in M$ and $k \neq \top$, if $\mathit{level(m)} = k$ then $\mathit{level}(\mathit{parent}(\mathit{member}(TS))) = k - 1$; \emph{(v)} $\mathit{parent}(\top) = \top$; \emph{(vi)} $\mathit{level}(\top) = 0$.

A time series belongs to a dimension's lowest level. Each member (except $\top$) at a level $k$ has a parent at level $k-1$. Users can analyze data at different levels by grouping on a level. To better describe the relation of the time series to real-world entities we use named levels. For example, the location dimension for the time series in our running example is \emph{Turbine $\rightarrow$ Park $\rightarrow$ Region $\rightarrow$ Country $\rightarrow \top$}. For a time series $TS$, $\mathit{member}(TS)$ returns a member for the \emph{Turbine} level, while  $\mathit{parent}(\mathit{member}(TS))$ returns a member for the \emph{Park} level. If a $TS$ is from a wind turbine with id $9834$ located in Aalborg, $\mathit{member}(TS) = 9834$, while $\mathit{parent}(9834) = \mathit{Aalborg}$ until $\mathit{parent}$ returns $\top$ indicating the top of the hierarchy.

\textbf{Time Series Group:}
A \emph{time series group} is a set of regular time series, possibly with gaps, $TSG = \{TS_1, \ldots, TS_n\}$, where for $TS_i, TS_j \in TSG$ it holds that they have the same sampling interval $SI$ and that $t_{1i} \bmod SI = t_{1j} \bmod SI$ where $t_{1i}$ and $t_{1j}$ are the first timestamp of $TS_{i}$ and $TS_{j}$, respectively.

For example, $TSG = \{TS, TS_{rg}\}$ is a time series group that contains the time series $TS$ and $TS_{rg}$, both with $SI = 100$ ms. The irregular time series $TS_g$ cannot be in the group as its $SI$ is undefined. Time series can be correlated even if they have very different values, as scaling can be applied.

So models can represent a time series group within $\epsilon$, the group can be split into dynamically sized sub-sequences. These segments use models to represent the values of the time series in the group, and each segment can use a different model type.

\begin{figure*} \centering
  \subcaptionbox{Grouping and partitioning on the master\label{fig:master}}{\includegraphics[width=0.332\textwidth]{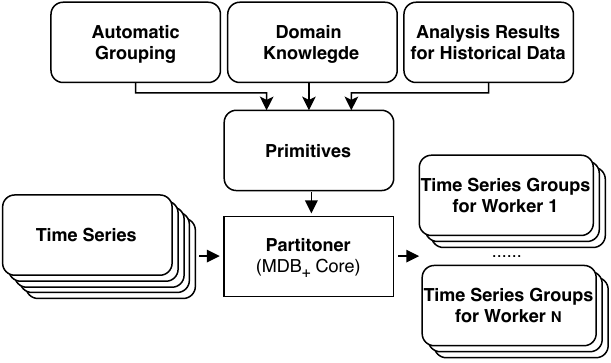}}
  \subcaptionbox{Architecture of \sysg{}'s worker nodes ingesting time series directly from their source\label{fig:worker}}{\includegraphics[width=0.662\textwidth]{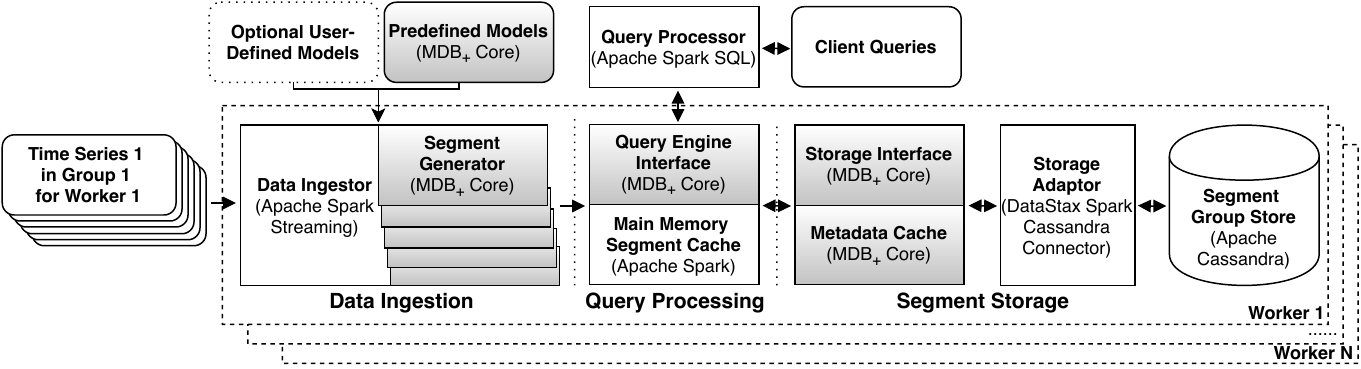}}
  \caption{\sysg{} groups and partitions time series on the master before the workers start ingesting the time series in parallel}\label{fig:architecture}
\end{figure*}

\textbf{Segment:}\label{def:segment}
A \emph{segment} is a 5-tuple $S = (t_s, t_e, SI, G_{ts} : TSG \to 2^{\{t_s, t_s+SI, \ldots, t_e\}}, m)$ representing the data points for a bounded time interval of a time series group $TSG$. The 5-tuple consists of start time $t_s$, end time $t_e$, sampling interval $SI$, a function $G_{ts}$ which for the $TS \in TSG$ gives the set of timestamps for which $v = \bot$ in $TS$, and where the values of all other timestamps are defined by the model $m$ multiplied by a scaling constant $C_{TS} \in \mathbb{R}$.

To exemplify segments we use three time series from collocated turbines: $TS_1 = \langle (100, 9.43),(200, 9.09), (300,$ $8.96), (400, 8.62), (500, 8.50) \rangle$, $TS_2 = \langle (100, 8.78), (200,$ $8.55), (300, 8.32), (400, 8.09), (500, 8.96) \rangle$, and $TS_3 = \langle (100,$ $9.49), (200, 9.20), (300, 8.92), (400, 8.73), (500, 8.65) \rangle$. Representing these with the linear function $m = -0.003t_i + 9.40$ creates an approximation with the error $|8.96 - (-0.003 \times 500 + 9.40)| = 1.06$ when using the uniform norm. If the error bound, e.g., is $1$, the segment $S = (100, 400, 100, G_{ts} = \emptyset, m = -0.003t_i + 9.40)$, $1 \leq i \leq 4$, is created.

 \section{Supporting \algo{} in \sysg{}}\label{sec:arch}
\subsection{Architecture of \sysg{}}\label{sec:arch:arch}
\sys{}~\cite{db:2018:modelardb} interfaces a library (\core{}) with the stock versions of Apache Spark for query processing and Apache Cassandra for storage in a master/worker architecture~\cite{db:2018:modelardb}. \sysg{} adds a \emph{Partitioner} to the master and updates \emph{all} of \sys{}'s components. The Partitioner exposes primitives for effectively specifying time series groups and an automatic grouping method as shown in Figure~\ref{fig:master} and detailed in Section~\ref{sec:partitioning}. When ingesting, \sysg{} compresses time series groups using \algo{} which selects an appropriate model type for each segment as described in Section~\ref{sec:rog}. Three model types, extended to compress groups as described in Section~\ref{sec:compression}, are part of \coreg{}: \ac{pmc}~\cite{model:2003:pmc}, \ac{swing}~\cite{model:2009:ss}, and \ac{fb}~\cite{models:2015:gorilla}.
A key benefit of \ac{pmc} and \ac{swing} is that they only use 32 bits and 64 bits per segment, respectively, regardless of the segment's length. But \ac{pmc} requires that the values can be approximated by a constant, while \ac{swing} requires that the values can be approximated by a linear function. This limitation is seen in Figure~\ref{fig:motivation} under \ac{mgc} where \ac{pmc} creates many segments to represent approximately linear time series. As \ac{fb} uses delta-of-delta encoding, the values need not follow a specific pattern, but \ac{fb} uses 1-32 bits per value. So \ac{pmc} and \ac{swing} are more efficient than \ac{fb} when time series follow constant or linear patterns, but can be inefficient if not. Users can also add more model types to \sysg{} without recompiling it. So by using multiple models types, \sysg{} can adapt as time series change at run-time.
A fallback model type that stores raw values is also provided. It is only used for segments that cannot be represented by any other model type within the user-defined bounds.

A worker consists of three sets of components as shown in Figure~\ref{fig:worker}. Each component is annotated with the software providing that functionality and components modified for \sysg{} are gray. \emph{Data Ingestion} constructs models of time series groups within user-defined bounds; \emph{Query Processing} caches segments and processes queries; \emph{Segment Storage} is a uniform interface with predicate push-down for the segment group store.
While adding our new \sysg{} methods to \sys{}, we also optimized \sys{}'s code as listed in Table~\ref{table:enhancements}.
Thus, \sysg{} builds upon \sys{} by adding our methods and optimizations.

\subsection{Ingestion and Gaps in \algo{}}\label{sec:rog}
\begin{figure}
  \centering
  \includegraphics[width=\columnwidth]{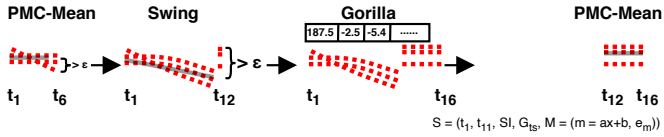}
  \caption{Multi-model ingestion using \algo{}}\label{fig:multi:ingestion}
\end{figure}

\algo{} consists of online ingestion, management of gaps, and dynamic splitting and merging of groups. For ingestion, \algo{} uses a window-based approach to support bounded and unbounded times series. Data points for a group are added to a window and a model (from a user-configurable list of model types) is fitted to all data points in the window at each SI. When a data point is received that the models cannot be fitted to within the error bound $\epsilon$, the model with the best compression is emitted. An example is shown in Figure~\ref{fig:multi:ingestion}. At $t_1$ a data point from each time series in the group is appended to a buffer. A model of the first model type (\ac{pmc}) is fitted to them. This is repeated for each $SI$ while possible within $\epsilon$. The model types in \coreg{} incrementally update models with new data points. At $t_6$, \ac{pmc} cannot represent all buffered data points within $\epsilon$. Therefore, \algo{} uses the next model type (\ac{swing}). If this model type can fit a model to all buffered data points ingestion continues, otherwise, the next model type is used. Lossless model types, e.g., \ac{fb}, are limited by a user-configurable length instead of $\epsilon$. At $t_{16}$ a data point is received that the last model type (\ac{fb}) cannot fit a model to, so \algo{} emits a segment with the model providing the best compression (a model of type Swing here). The data points the model represents are removed from the buffer and ingestion restarted using the first model type (\ac{pmc}). \algo{} creates disconnected segments (no duplicate data points) for better compression compared to connected segments~\cite{rw:multi:2011:towards}. If the model types in the list cannot represent any of the buffered data points, the fallback model type is used, see Section~\ref{sec:arch:arch}.

\begin{table}
\caption{Optimizations Implemented in \sysg{}}
\label{table:enhancements}
\resizebox{\columnwidth}{!}{
\begin{tabular}{l p{2.7in}}
        \toprule
        Component Set & Change \\
        \midrule
        Data Ingestion   & Read files in smaller chunks to reduce memory usage.\\ & Write segments directly to storage when bulk-loading.\\ & Replace PMC-MR with PMC-Mean to lower average error.\\ & Pre-allocate memory for Gorilla based on the length bound.\\ Query Processing & Use static and dynamic code generation for faster projections.\\ & Use the number of Spark partitions as a caching heuristic.\\ & Parallelize queries over Gids in Spark instead of Cassandra.\\ Segment Storage  & Only decompress segments when used for query processing.\\ \bottomrule
\end{tabular}}
\end{table}

\begin{figure}
  \centering
  \includegraphics[width=0.67\columnwidth]{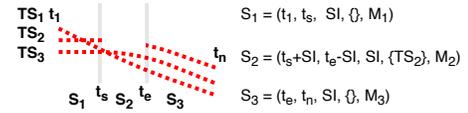}
  \caption{Removing gaps by creating new segments}\label{fig:multi:gaps}
\end{figure}

If a gap occurs \algo{} creates a new segment as shown in Figure~\ref{fig:multi:gaps}. A model is fitted to data points from all three time series at $t_1$. At $t_s + SI$ a gap occurs in $TS_2$ and a new model is fitted to the data points from only two time series. As this model represents a subset of the group, the ids of the time series not represented are stored in the segment, see $S_2$. When data points are received from $TS_2$ again, the process is repeated, see $t_e$ and $S_3$. Thus, a segment represents data points for a static number of time series. This simplifies implementing user-defined model types as \sysg{} manages gaps.

\subsection{Storage Schema}\label{sec:schema}
\begin{figure}
  \centering
  \includegraphics[width=\columnwidth]{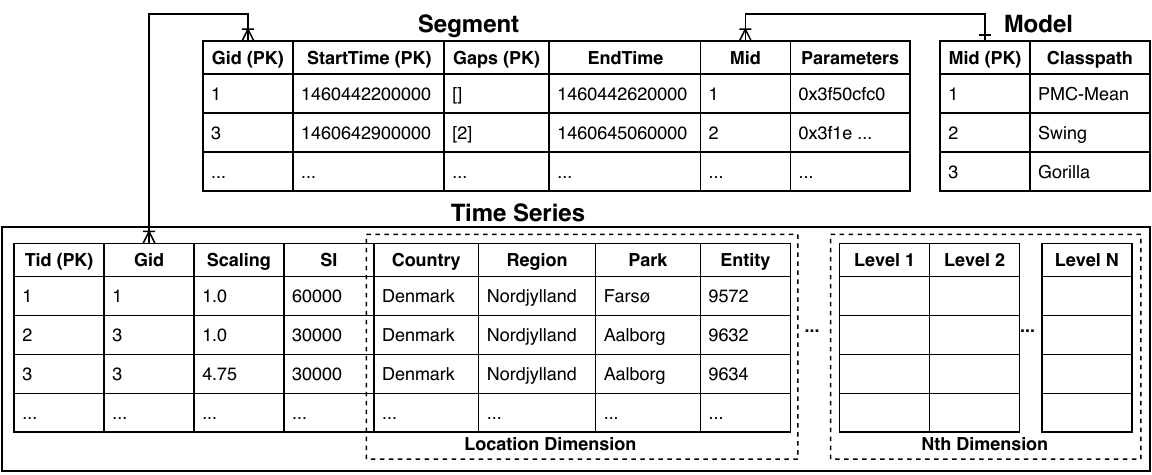}
  \caption{General Schema for storing groups as segments}\label{fig:schema}
\end{figure}

The general schema used by \sysg{} to support \algo{} is shown in Figure~\ref{fig:schema}. \texttt{Time Series} stores \texttt{SI}, a \texttt{Scaling} constant ($C_{TS}$ in the definition of Segment), a group id \texttt{gid}, and user-defined denormalized dimensions for each time series, with each identified by a \texttt{Tid}. Only the \texttt{SI} must be set by the user. \texttt{Gid} is the group a time series has been assigned to by \sysg{} using the provided primitives. \texttt{Scaling} defaults to $1.0$ and is applied to values during ingestion and query processing. \texttt{Model} maps a \texttt{Mid} to the Java Classpath of that model type. \texttt{Segment} stores data points as dynamically sized segments.
\texttt{Segment}'s primary key includes \texttt{Gaps} to prevent duplicate keys due to the dynamic splitting described in Section~\ref{sec:partitioning:dynamic}. While user-defined dimensions are stored in \texttt{Time Series}, no explicit time dimension is needed as aggregates in the time dimension can be computed efficiently using only \texttt{StartTime} and \texttt{EndTime} as described in Section~\ref{sec:query:time}.
By exploiting that the regular time series with gaps in each group are aligned, their timestamps can be efficiently stored as \texttt{StartTime}, \texttt{EndTime}, \texttt{SI} in \texttt{Time Series}, and \texttt{Gaps}.
\texttt{Segment} is modified for Cassandra. To improve predicate push-down, its primary key is changed to \texttt{Gid, EndTime, Gaps}~\cite{db:2018:modelardb}. \texttt{Gaps} are 64-bit integers with bits set to $1$ if that time series in the group has a gap. Also, the column \texttt{StartTime} is stored as the segment's size to save space, but can recomputed as \texttt{StartTime = EndTime - (Size $\times$ SI)}~\cite{db:2018:modelardb}.
 \section{Grouping and Partitioning}\label{sec:partitioning}
\subsection{Motivation}\label{sec:partitioning:motivation}
To provide the benefit of model-based storage and query processing while ensuring low latency, models must be fitted online~\cite{db:2018:modelardb}. Also, in a distributed system, time series in a group should be ingested by one node to minimize latency and bandwidth use. Thus, the time series must be grouped and partitioned before ingestion begins based on, e.g., metadata, historical data, or domain knowledge. As the best set of groups depends on the model types used, \sysg{} has primitives so users effectively can specify which time series to group. Users can thus use their preferred grouping method with \sysg{}. Locating correlation through data mining~\cite{correlation:2018:boehm:journal, correlation:2020:edbt} is thus an orthogonal problem. \sysg{} also has automatic grouping using the time series dimensions. Thus, this can be used when historical data or enough resources to analyze it are not available.

\subsection{Primitives and Automatic Grouping}\label{sec:partitioning:primitives}
\sysg{}'s primitives allow groups to be specified as sets of time series, members that must be equal, levels for which members must be equal, or the \emph{distance} between the dimensions (see below). They can be combined using \texttt{AND} and \texttt{OR}.

 \begin{figure}
   \centering
   \includegraphics[width=0.8\columnwidth]{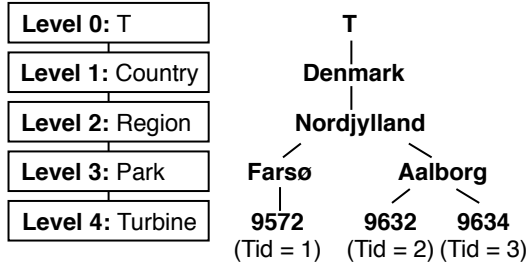}
   \caption{The location dimension for (a subset of) our running example. The LCA for $Tid = 2$ and $Tid = 3$ is the level \emph{Park}}\label{fig:lca}
\end{figure}

To group specific time series, their source (file or socket) must be provided, e.g., \texttt{4aTemp.gz 4bTemp.gz}. A scaling constant can be added per time series. This provides precise control but is time-consuming. The other primitives allow time series to be grouped based on their dimensions, e.g., temperature sensors in proximity likely produce similar values. The similarity of dimensions for two groups can be computed as their \ac{lca} level. This is the lowest level in a dimension where all time series in the groups share members starting from $\top$. An example is shown in Figure~\ref{fig:lca}.

A group based on members is specified as a triple with a dimension, a level, and a member or a pair with a dimension and an \ac{lca} level. The triple \texttt{Measure 1 Temperature}, e.g., groups time series with the member \texttt{Temperature} at level one of the \texttt{Measure} dimension. The pair \texttt{Location 2} groups time series if their \ac{lca} level is equal to or higher than two for the \texttt{Location} dimension. Zero specifies that all levels must be equal, and a negative number $n$ that all but the lowest $|n|$ levels must equal. A scaling constant can be defined for time series with a specific member as a 4-tuple of dimension, level, member, and constant. These primitives are intended for data sets with few dimensions but many series.

For data sets with many time series and dimensions, users can specify groups by the \emph{distance} $[0.0;1.0]$ between dimensions. Intuitively, time series with high overlap between their members are correlated. The time series in Figure~\ref{fig:lca} are, e.g., more likely correlated if they share members at the Turbine level than the Country level. For the distance $0.0$ all members must match for time series to be grouped, and for $1.0$ all time series are grouped. Values in-between specify different degrees of overlap.

To \textit{automatically} group time series, users can specify correlation as \texttt{auto}, making \sysg{} group time series within the lowest non-zero distance possible in the data set. This distance is given by $(1.0 / \mathit{\max(Levels)}) / \mathit{|\mathbb{D}|}$ where $\mathit{Levels}$ is the set of levels in each dimension and $\mathbb{D}$ is the dimensions.

\subsection{Static Grouping and Partitioning}\label{sec:partitioning:partitioner}
\begin{figure}
  \centering
  \includegraphics[width=\columnwidth]{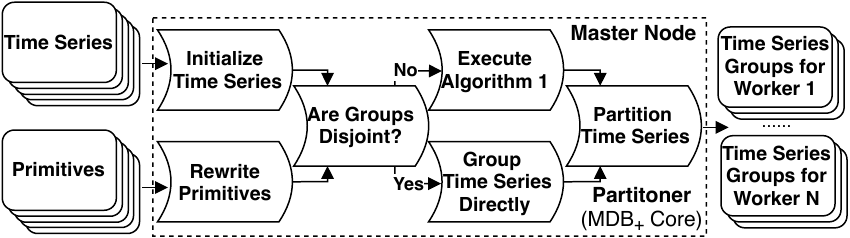}
  \caption{Offline grouping and partitioning of time series}\label{fig:architecture:partitioner}
\end{figure}

\sysg{}'s Partitioner is shown in Figure~\ref{fig:architecture:partitioner}. Based on the user's configuration (time series, primitives), the Partitioner groups and partitions the time series across the cluster. For each time series, a reader is created based on the time series source. All references to levels in the dimensions are rewritten to indexes for \sysg{}'s denormalized schema shown in Section~\ref{sec:schema}, and \texttt{auto} is rewritten to the shortest non-zero distance.

Groups are constructed from the rewritten primitives. First, \sysg{} tries to determine if the primitives create disjoint groups, and if it can, adds each time series directly to its group, otherwise, Algorithm~\ref{algo:group} is used. In Line~1 a group is created per time series. Then, in Line~2--12, for each user-defined correlation $\mathit{clause}$, groups are merged until a fixed point is reached in the number of groups. In Line~9 $\mathit{correlated}$ evaluates the $\mathit{clause}$ and ensures that all time series in the groups are correlated before they are merged. As the clauses are applied in their defined order, users control their priority. In essence, Algorithm~\ref{algo:group} computes cliques (as correlations are non-transitive) in a graph with time series as vertices and $\mathit{correlated}$ defining the edges. However, as edges are dynamically defined as groups of correlated series are found, all possible edges need not be materialized. While its complexity is $O(K \times N^3)$ for $K$ clauses and $N$ time series, grouping is done \emph{once} before ingestion and the actual run-time is \emph{very low}, see Section~\ref{sec:evaluation}.

\begin{algorithm}[t]
  \SetKwInput{Input}{Input}
  \DontPrintSemicolon{}

  \Input{A list of time series $\mathbb{TS}$\newline
    The dimensions for all time series $\mathbb{D}$\newline
    A list of correlation clauses $\mathit{Correlations}$
  }
  $\mathit{TSG} \gets \mathit{createSingleTimeSeriesGroups(\mathbb{TS})}$\;
  \ForEach{$\mathit{clause} \in \mathit{Correlations}$} {$\mathit{groupsModified} \gets \mathit{true}$\;
    \While{$\mathit{groupsModified}$} {$\mathit{groupsModified} \gets \mathit{false}$\;
      \ForEach{$i \gets 1 \textbf{ to } \mathit{|TSG|}$} {\ForEach{$j \gets i + 1 \textbf{ to } \mathit{|TSG|}$} {$\mathit{(TSG_1, TSG_2)} \gets \mathit{getPairAt(TSG, i, j)}$\;
          \If{$\mathit{correlated(clause, TSG_1, TSG_2, \mathbb{D})}$} {$\mathit{TSG[i]}\gets{} mergeGroups(TSG_1, TSG_2)$\;
            $removeGroupAfterLoop(TSG_2, TSG)$\;
            $\mathit{groupsModified} \gets{} \mathit{true}$\;
          }
        }
      }
    }
  }
  \KwRet{$\mathit{Groups}$}\;\caption{Group series using the primitives}\label{algo:group}
\end{algorithm}

To evaluate the primitives, $\mathit{correlated}$ uses different functions. We only describe the one grouping by distance, as the remaining follow directly from the description of the primitives. First, the distance between the dimensions of the two time series groups is calculated as $dist = (\sum_{d \in \mathbb{D}} \mathit{weight_d} \times ((\mathit{levels_d} - \mathit{lca_d}) / \mathit{levels_d})) / |\mathbb{D}|$ where $\mathit{levels_d}$ is the number of levels in $d$, and $\mathit{lca_d}$ is the two groups' LCA level for $d$.
Users can change the impact of a dimension by a $\mathit{weight}$ (default is $1$).
As it is more intuitive to increase the weight for important dimensions than to decrease it, $weight_d$ is the reciprocal of the user-provided weight.
For the dimension in Figure~\ref{fig:lca}, the distance between the time series with $Tid = 2$ and $Tid = 3$ is, e.g., $1 \times ((4 - 3) / 4) = 0.25$. The distance is then capped by $\min(\mathit{dist}, 1.0)$, and the groups merged if the distance is at most the threshold set by the user or \texttt{auto}.

Last, a set of time series groups $\mathbb{TSG}$ is created so each worker receives approximately the same number of data points per second. The method used is based on~\cite{system:2009:multi}, and tries to minimize $\max_{SG_1 \in \mathbb{TSG}}(data\_points\_per\_minute(SG_1)) - \min_{SG_2 \in \mathbb{TSG}}(data\_points\_per\_minute(SG_2))$.

\subsection{Dynamically Splitting Groups}\label{sec:partitioning:dynamic}
\begin{figure}
  \centering
  \includegraphics[width=0.8\columnwidth]{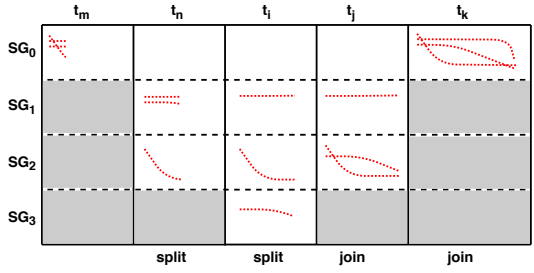}
  \caption{Dynamic splitting and merging in \algo{}}\label{fig:sj}
\end{figure}

As events can change the values of a time series, e.g., a turbine can be damaged, \algo{} splits a group if its time series become temporarily uncorrelated. This is shown in Figure~\ref{fig:sj}. \algo{} discards emitted data points, but they are included in the example to show how the time series change over time. At $t_{m}$ the group is ingested using the Segment Generator $SG_{0}$. At $t_{n}$ its time series are no longer correlated leading to poor compression. The group is then split into two and ingestion continues with $SG_{1}$ and $SG_{2}$. $SG_{0}$ synchronizes ingestion to simplify merging and merges the split groups if they become correlated again. Then at $t_{i}$ the group ingested by $SG_1$ is split causing each time series to be ingested separately.

\begin{algorithm}
  \SetKwInput{Input}{Input}
  \DontPrintSemicolon{}

  \Input{A time series group $\mathit{TSG}$\newline
    The user-defined error bound $\mathit{\epsilon}$\newline
    The data points buffered for $TSG$ in $\mathit{Buffer}$
  }
  $\mathit{Splits} \gets \mathit{createSet()}$\;
  \While{$\mathit{notEmpty(TSG)}$} {$\mathit{TS_1} \gets \mathit{getTimeSeries(TSG)}$\;
    $\mathit{TSG_{n}} \gets \mathit{createGroup()}$\;
    \ForEach{$\mathit{TS_2} \in \mathit{TSG}$} {$\mathit{DP_1} \gets \mathit{dataPoints(TS_{1}, Buffer)}$\;
      $\mathit{DP_2} \gets \mathit{dataPoints(TS_{2}, Buffer)}$\;
      \If{$\mathit{allWithinDoubleBound(DP_1, DP_2, \epsilon)} $} {$\mathit{addTimeSeriesToGroup(TS_2, TSG_{n})}$\;
        $\mathit{removeTimeSeriesFromGroup(TS_2, TSG)}$\;
      }
    }
    $\mathit{sg} \gets createSegmentGenerators(TSG_{n})$\;
    $\mathit{addSegmentGeneratorToSet(sg, Splits)}$\;
  }
  \KwRet{$\mathit{Splits}$}\;\caption{Potentially splitting $TSG$ temporarily}\label{algo:split}
\end{algorithm}

To reduce the number of non-beneficial splits and the overhead of determining when to split, \algo{} uses two heuristics: poor compression ratio and the error between buffered data points. If the compression ratio of a new segment is below a user-configurable fraction of the average (default is $1/10$) and data points are buffered, Algorithm~\ref{algo:split} is run. A new segment indicates that the structure of the time series has changed as the next value would exceed the error or length bound. Algorithm~\ref{algo:split} groups time series if their buffered data points are within twice the user-defined error bound ($2\epsilon$) using the error function of the last emitted model in Line~6--10. Thus, groups of size one to $\mathit{|TSG|}$ can be created. In Figure~\ref{fig:sj} two groups are created both at $t_n$ ($SG_1$ and $SG_2$) and at $t_i$ ($SG_1$ and $SG_3$). Time series currently in a gap are grouped.

At $t_{j}$ in Figure~\ref{fig:sj} two of the time series become correlated again and are merged into one group. Then, at $t_{k}$ all the time series are correlated again so $SG_{0}$ takes over ingestion. Like for splitting, \algo{} merges groups by grouping time series if their buffered data points are within $2\epsilon$. However, when merging, only one time series from each group is compared as the groups consist of correlated time series (if not a split would have occurred). To simplify merging, it is only attempted at the end of a $SI$ when all groups have received data points for the period. To reduce the overhead of merging, a split group is only marked for merging after a number of segments are emitted. The required number of segments starts at one and is doubled after each merge attempt. The intuition is that a failed merge attempt further indicates that the splits are preferable.

 \begin{figure}
  \centering
  \includegraphics[width=0.96\columnwidth]{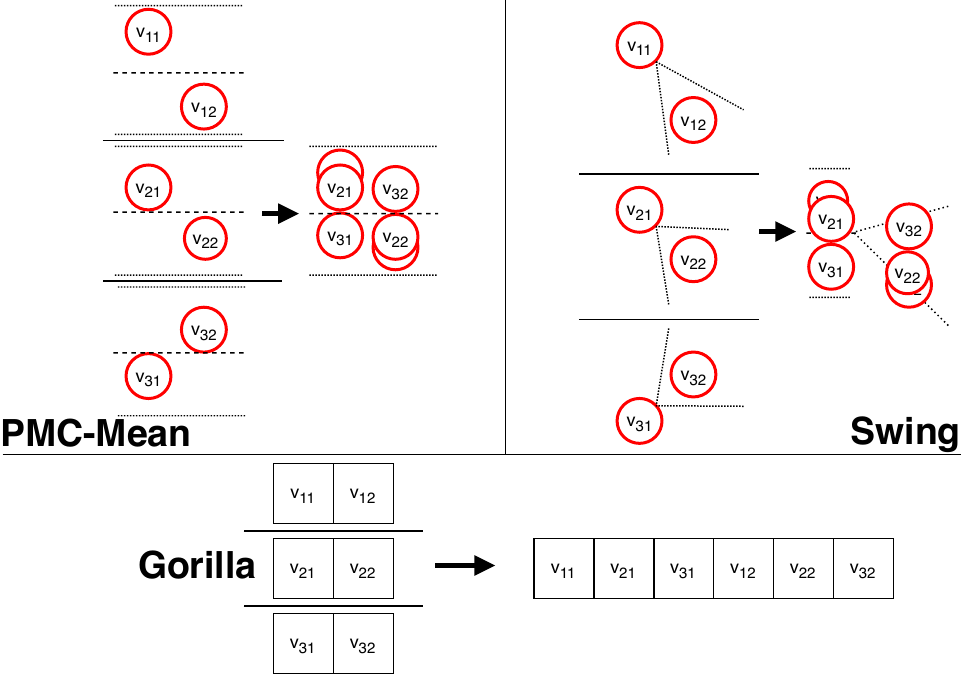}
  \caption{Extending model types to support groups. For lossy types the x-axis is time and the y-axis is value. Full lines are separators, dashed are the average value, and dotted show $\epsilon$}\label{fig:translation}
\end{figure}

\section{Model Extensions}\label{sec:compression}
\ac{mmgc} needs multiple model types that can fit models to time series groups. However, most existing model types fit models to only one time series~\cite{rw:2013:model:survey, rw:2013:model:evaluation}. A segment can store a model per time series, but it only reduces the amount of metadata. Instead, \coreg{} uses model types extended to fit models to a \emph{time series group} based on two general ideas.

For model types that fit models to values by ensuring they are within an upper and lower bound according to the uniform norm, e.g., \ac{pmc} and \ac{swing}, we only need to ensure that the minimum and maximum value for each timestamp are within the error bound. \ac{pmc} represents a set of values $V \subset \mathbb{R}$ as $\avg(V)$ within $\epsilon$ of $\min(V)$ and $\max(V)$. Thus, \ac{pmc} needs no changes as it only tracks $\min(V)$, $\max(V)$, and $\avg(V)$. See \ac{pmc} in Figure~\ref{fig:translation}. \ac{swing} produces a linear function that intersects with the first value and can represent subsequent values within $\epsilon$. For groups, the first value can be computed using \ac{pmc}. The following values are added one at a time. See \ac{swing} in Figure~\ref{fig:translation}.

For model types using lossless compression, e.g., \ac{fb}, values from a time series group should be stored in time ordered blocks. This allows exploitation of both temporal correlation and correlation across time series. As the time series in a group are correlated, $n-1$ values in a block have a small delta-of-delta compared to the previous value, which \ac{fb} encodes using only a few bits per value. See \ac{fb} in Figure~\ref{fig:translation}.

\begin{figure*}
  \centering
  \includegraphics[width=\textwidth]{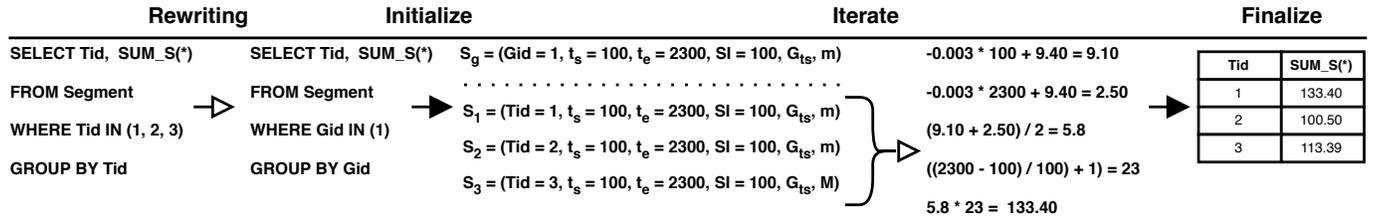}
  \caption{Aggregation performed on the linear model $-0.0465t + 186.1$ representing a time series group}\label{fig:query:simple}
\end{figure*}

As a glimpse of the benefit of our extensions, we compress seven real-life time series together. These time series contain measurements of energy frequency from a small park of wind turbines. By grouping them, \sysg{} uses 67.2\% less storage ($\epsilon = 0\%$) compared to compressing them separately.
 \section{Query Processing}\label{sec:query}
\subsection{Query Interface}
As a model can reconstruct its data points within $\epsilon$, \sysg{} supports arbitrary SQL queries on data points using a Data Point View with the schema \texttt{(Tid int, TS timestamp, Value float, <Dimensions>)}.\,\texttt{<Dimensions>} are the denormalized user-defined dimensions. They are cached in-memory and added during query processing. Some models (e.g., \ac{pmc} and \ac{swing}) can compute many aggregates in constant time~\cite{db:2018:modelardb}. Thus, aggregates can be computed in linear time in the number of models instead of the number of data points. As a model usually represents many data points, this greatly reduces query time as shown in Section~\ref{sec:evaluation}. Aggregates on models are provided as UDAFs on a Segment View with the schema \texttt{(Tid int, StartTime timestamp, EndTime timestamp, SI int, Mid int, Parameters blob, Gaps blob, <Dimensions>)}. UDAFs for simple aggregates are suffixed with \texttt{\_S}, e.g., \texttt{MAX\_S}, while UDAFs for aggregates in the time dimension are named \textit{CUBE\_<AGGREGATE>\_<INTERVAL>}, e.g., \texttt{CUBE\_AVG\_HOUR}. Aggregates on models in the user-defined dimensions can be reduced to simple aggregates with a \texttt{GROUP BY} on the appropriate columns in the Segment View. As a result, we only show how simple aggregates and aggregates in the time dimension can be executed on models.

\subsection{Aggregate Queries}\label{sec:query:simple}
To allow users to query time series instead of time series groups, a mapping between $Tid$s and $Gid$s is performed during query processing using the ids stored in the \texttt{Time Series} table, see Figure~\ref{fig:schema}. Queries and results only use $Tid$s. $Gid$s are pushed to the segment group store so it only has to index $Gid$s. \sys{} only supports predicate push-down for \texttt{Tid}, \texttt{StartTime}, and \texttt{EndTime}~\cite{db:2018:modelardb}, but \sysg{} also pushes user-defined dimensions by rewriting members in the \texttt{WHERE} clause to the $Gid$s of the groups with time series with these members.

A simple aggregate using the Segment View is shown in Figure~\ref{fig:query:simple}. First, the master replaces the $Tid$s and members in the queries' \texttt{WHERE} clause with $Gid$s and sends it to the workers (Rewriting). Then, each worker creates an iterator for the relevant segments in its segment group store (Initialize) and computes the aggregate for each segment read using the iterator (Iterate). Last, to support distributive and algebraic functions, the result is computed from the intermediates (Finalize).

\begin{figure*}
  \centering
  \includegraphics[width=\textwidth]{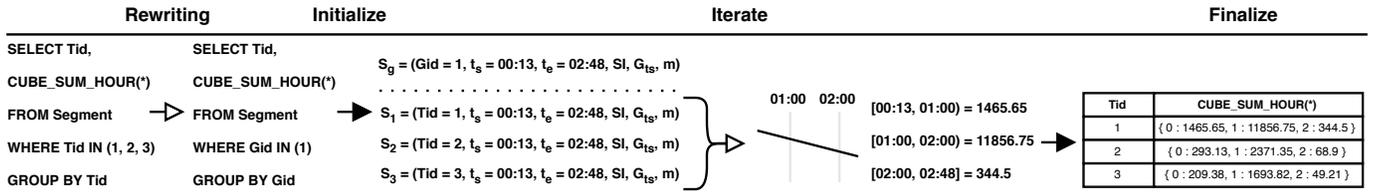}
  \caption{Aggregation in the time dimension on a linear model representing a group of three time series}\label{fig:query:time}
\end{figure*}

\subsection{Aggregate Queries in the Time Dimension}\label{sec:query:time}
As the \texttt{Segment} table stores the start time and end time for each segment, see Figure~\ref{fig:schema}, aggregates in the time dimension can be computed from it. An example of aggregation in the time dimension using the Segment View is shown in Figure~\ref{fig:query:time}. The query computes the sum per hour on models representing the time series with $Tid=1$--$3$ using the UDAF \texttt{CUBE\_SUM\_HOUR}. After rewriting, the aggregate is computed for the interval from $t_s = 00{:}13$ until $01{:}00$ which is the next time\-stamp delimiting two aggregation intervals. Then it is computed from $01{:}00$ until $02{:}00$, and from $02{:}00$ to and including $t_e = 02{:}48$. The last interval is inclusive as the segments are disconnected, see Section~\ref{sec:rog}. The pseudo-code for executing aggregates in the time dimensions using the Segment View is shown in Algorithm~\ref{algo:at}. In Line 1 the query is rewritten (Rewriting), each worker creates an iterator for the relevant segments in Line 2 (Initialize), and in Line 3--10 the aggregate is computed for each segment and time interval (Iterate). Last, the final result is computed and returned in Line 13 to support distributive and algebraic functions (Finalize).

\begin{algorithm}
  \SetKwInput{Input}{Input}
  \DontPrintSemicolon{}
  \SetNoFillComment{}

  \Input{The \texttt{WHERE} clause of the query $\mathit{predicates}$\newline
    Mapping from Gid to Tid and reverse $\mathit{groups}$\newline
    mapping from Members to Gid $\mathit{members}$\newline
    The roll-up level in the time dimension $\mathit{level}$\newline
    A segment aggregation function $\mathit{iterate}$\newline
    A function for aggregating results $\mathit{finalize}$
  }

  $\mathit{predicates} \gets \mathit{rewrite(predicates, groups, members)}$\;
  \tcc{The rewritten query is run on the workers in parallel}
  $\mathit{segments} \gets{} \mathit{getSegmentIterator(predicates, groups)}$\;
  \ForEach{$\mathit{segment} \in \mathit{segments}$} {$\mathit{originalEndTime} \gets{} \mathit{extractEndTime(segment)}$\;
    $\mathit{startTime} \gets{} \mathit{extractStartTime(segment)}$ \;
    $\mathit{endTime} \gets{} \mathit{ceilToLevel(startTime, level)}$\;
\While{$\mathit{endTime} < \mathit{originalEndTime}$} {$\mathit{result} \gets{} \mathit{aggregateInterval(iterate, segment,}$\\\nonl\ \hspace{1.23cm}$\mathit{startTime, endTime, result)}$\;
        $\mathit{startTime} \gets{} \mathit{endTime}$\;
        $\mathit{endTime} \gets{} \mathit{updateForLevel(startTime, level)}$\;
      }
$\mathit{result} \gets{} \mathit{aggregateInterval(iterate, segment,}$\\\nonl\ \hspace{1.23cm}$\mathit{startTime, originalEndTime, result)}$\;
    }
  $\mathit{results} \gets \mathit{mergeResults()}$\;
  \KwRet{} $\mathit{finalize(results)}$\;
  \caption{Execution of aggregate queries with a roll-up in the time dimension on the Segment View}\label{algo:at}
\end{algorithm}
 \section{Evaluation}\label{sec:evaluation}
\subsection{Overview and Evaluation Environment}\label{sec:overview}
We compare \sysg{} to state-of-the-art big data formats and systems used in industry (ORC, Parquet, Cassandra), the popular TSMS InfluxDB (together termed industry formats), and the model-based \ac{mmc} \ac{tsms} \sys{}~\cite{db:2018:modelardb}. The industry formats all use lossless compression, while \sys{} and \sysg{} are model-based and support lossless and lossy compression. RDBMSs are not included due to their poor compression for time series~\cite{db:2018:modelardb}. To separate the effect of our methods from the difference in implementation between \sys{} and \sysg{}, we evaluate \sysg{} using the \emph{best} correlation primitives (\sysgonb{}), using \texttt{auto} with weighted dimensions (\sysgona{}), and with grouping disabled (\sysgoff{}). \sysgoff{} should be faster when queries only require a few time series per group, e.g., simple aggregate and point/range queries, while \sysgona{} and \sysgonb{} should perform better for ingestion, compression, and queries on entire groups. \sysg{} can thus be configured depending on the exact use case. We use seven nodes, each with an i7-2620M, 8 GiB of 1333 MHz DDR3 memory, a 7,200 RPM hard drive, and 1 Gbit Ethernet. Ubuntu 16.04, InfluxDB 1.4.2, Hadoop 2.8.0, Spark 2.1.0, and Cassandra 3.9 are installed on ext4. The \emph{master} is a \emph{Primary / Secondary HDFS NameNode} and \emph{Spark Master}, while the six \emph{worker}s are \emph{Cassandra Node}s, \emph{HDFS Datanode}s, and \emph{Spark Slave}s. Spark is used for query processing, except for InfluxDB as the open-source version and existing Spark connectors do not support distribution. Instead, we partition each data set across all workers using range-partitioning and execute queries across workers in parallel from the master. Default configurations are used when possible, and changed values are selected to work well with the hardware and data sets used. For \sysg{} we use length limit $50$, split fraction $10$, and write segments in batches of $50{,}000$. The default error bound is 10\% for \sys{} and \sysg{}. InfluxDB-Java 2.10 is used, and data points are written in batches of $50{,}000$. Spark is limited to 4 GiB driver memory and 3 GiB executor memory so Cassandra and HDFS are stable, predicate push-down is enabled for all formats, and automatic unpersist of micro-batches is disabled. For Cassandra, the commitlog is set to 128 MiB, maximum batch size to 50 MiB, and DataStax Spark Connector version 2.0.3 is used. We only run the required software and disable replication.
Graphs show the number of workers on the x-axis labels, use striped bars for \sysgonb{} and \sysgona{}, and have a gap on the y-axis if the results differ very significantly between formats.
\sysg{} is designed for unbounded time series, but our evaluation uses three static real-life data sets. Unbounded time series are created by repeating static time series. The industry formats use the schema: \texttt{(Tid int, TS timestamp, Value float, <Dimensions>)}. \texttt{timestamp} is each format's timestamp type. Cassandra uses \texttt{(Tid, TS, Value)} as primary key, InfluxDB stores time series as one measurement with \texttt{Tid} as a tag, and for ORC and Parquet we store a file per series on HDFS in a folder named \texttt{Tid=n} so Spark can prune by \texttt{Tid}.

\subsection{Data Sets and Queries}\label{sec:data:queries}
\textbf{Data Set ``EP''}
This real-life data set consists of 45,353 regular time series with gaps from energy production (e.g., humidity and wind speed) for 508 days, has $SI = 60$ s, has two dimensions \emph{Production: Entity $\rightarrow$ Type} and \emph{Measure: Concrete $\rightarrow$ Category}, and uses 339 GiB as uncompressed CSV.

\textbf{Data Set ``EF''}
This real-life data set consists of 197 regular time series with gaps from wind parks (e.g., rotation speed and temperature). The data was collected with an approximate $SI = 100$ ms from a Windows OPC DA server. A pre-processing step rounds timestamps to the nearest 200 ms and removes data points with equal time\-stamps (due to collection limitations not present in a production setup). It has two dimensions \emph{Location: Entity $\rightarrow$ Park $\rightarrow$ Country} and \emph{Measure: Concrete $\rightarrow$ Category},  and uses 372 GiB as uncompressed CSV.

\textbf{Data Set ``HD''}
\sysg{} was motivated by the energy domain, but we demonstrate its generality using a data set from \href{http://histdata.com}{histdata.com} (e.g., currency exchange rates, commodities, and indexes). It consists of 330 regular time series with gaps from 2000-2019, has $SI = 60$ s, one dimension \emph{Forex: Concrete $\rightarrow$ Category $\rightarrow$ Pair},  and uses 32.75 GiB as uncompressed CSV.

\textbf{Queries}
The queries are based on discussions with turbine owners: small simple aggregate queries simulate interactive analysis (S-AGG), large-scale simple aggregate queries evaluate scalability (L-AGG), medium-scale multi-dimensional aggregate queries simulate reporting (M-AGG). Point/range (P/R) queries extract sub-sequences. Half of S-AGG aggregate one time series while the rest \texttt{GROUP BY} \texttt{Tid} for five time series. L-AGG aggregate the full data set where half \texttt{GROUP BY} \texttt{Tid}. M-AGG is multi-dimensional aggregates on energy production measures. Half \texttt{GROUP BY} month and dimension, while the rest \texttt{GROUP BY} month, dimension and \texttt{Tid}. P/R queries (not \sysg{}'s intended use case but included for completeness) have \texttt{WHERE} clauses on either \texttt{TS} or \texttt{Tid} and \texttt{TS}.

\begin{figure*}
    \centering
    \begin{minipage}[b]{.24\textwidth}
        \resizebox{\columnwidth}{0.75\columnwidth}{\input{Figures/eval-ingestion.pgf}}
        \caption{Ingestion, EP}%
        \label{fig:query:ingestion}
    \end{minipage}
    \begin{minipage}[b]{.24\textwidth}
        \resizebox{\columnwidth}{0.75\columnwidth}{\input{Figures/eval-compression-ep.pgf}}
        \caption{Storage, EP}%
        \label{fig:error:ep}
    \end{minipage}
    \begin{minipage}[b]{.24\textwidth}
        \resizebox{\columnwidth}{0.75\columnwidth}{\input{Figures/eval-compression-eh.pgf}}
        \caption{Storage, EF}%
        \label{fig:error:eh}
    \end{minipage}
    \begin{minipage}[b]{.24\textwidth}
        \resizebox{\columnwidth}{0.75\columnwidth}{\input{Figures/eval-compression-hd.pgf}}
        \caption{Storage, HD}%
        \label{fig:error:hd}
    \end{minipage}
    \begin{minipage}[b]{.24\textwidth}
        \resizebox{\columnwidth}{0.75\columnwidth}{\input{Figures/eval-models-datapoints-ep.pgf}}
        \caption{Format, EP}%
        \label{fig:error:models:ep}
    \end{minipage}
    \begin{minipage}[b]{.24\textwidth}
        \resizebox{\columnwidth}{0.75\columnwidth}{\input{Figures/eval-models-datapoints-eh.pgf}}
        \caption{Format, EF}%
        \label{fig:error:models:eh}
    \end{minipage}
    \begin{minipage}[b]{.24\textwidth}
        \resizebox{\columnwidth}{0.75\columnwidth}{\input{Figures/eval-models-datapoints-hd.pgf}}
        \caption{Format, HD}%
        \label{fig:error:models:hd}
    \end{minipage}
    \begin{minipage}[b]{.24\textwidth}
        \resizebox{\columnwidth}{0.75\columnwidth}{\input{Figures/eval-distance.pgf}}
        \caption{Size/Distance}%
        \label{fig:error:dist}
    \end{minipage}
    \begin{minipage}[b]{.24\textwidth}
        \resizebox{\columnwidth}{0.75\columnwidth}{\input{Figures/eval-scalability-l-agg.pgf}}
        \caption{L-AGG, EP}%
        \label{fig:query:lagg}
    \end{minipage}
    \begin{minipage}[b]{.24\textwidth}
        \resizebox{\columnwidth}{0.75\columnwidth}{\input{Figures/eval-scalability.pgf}}
        \caption{Scale, L-AGG}%
        \label{fig:query:scale}
    \end{minipage}
    \begin{minipage}[b]{.24\textwidth}
        \resizebox{\columnwidth}{0.75\columnwidth}{\input{Figures/eval-query-s-agg-ep.pgf}}
        \caption{S-AGG, EP}%
        \label{fig:query:sagg:ep}
    \end{minipage}
    \begin{minipage}[b]{.24\textwidth}
        \resizebox{\columnwidth}{0.75\columnwidth}{\input{Figures/eval-query-s-agg-eh.pgf}}
        \caption{S-AGG, EF}%
        \label{fig:query:sagg:eh}
    \end{minipage}
    \begin{minipage}[b]{.24\textwidth}
        \resizebox{\columnwidth}{0.75\columnwidth}{\input{Figures/eval-query-m-agg-one-ep.pgf}}
        \caption{M-AGG-1, EP}%
        \label{fig:query:magg:ep:one}
    \end{minipage}
    \begin{minipage}[b]{.24\textwidth}
        \resizebox{\columnwidth}{0.75\columnwidth}{\input{Figures/eval-query-m-agg-two-ep.pgf}}
        \caption{M-AGG-2, EP}%
        \label{fig:query:magg:ep:two}
    \end{minipage}
    \begin{minipage}[b]{.24\textwidth}
        \resizebox{\columnwidth}{0.75\columnwidth}{\input{Figures/eval-query-m-agg-one-eh.pgf}}
        \caption{M-AGG-1, EF}%
        \label{fig:query:magg:eh:one}
    \end{minipage}
    \begin{minipage}[b]{.24\textwidth}
        \resizebox{\columnwidth}{0.75\columnwidth}{\input{Figures/eval-query-m-agg-two-eh.pgf}}
        \caption{M-AGG-2, EF}%
        \label{fig:query:magg:eh:two}
    \end{minipage}
    \hfill
\end{figure*}

\subsection{Experiments}\label{sec:evaluation:experiments}
\textbf{Ingestion Rate}\label{sec:evaluation:experiments:ingestion}
\sysg{}'s ingestion rate is evaluated by \emph{bulk loading} without queries (B) on one node to directly compare with InfluxDB. We ingest EP energy production measures ($3500$ gzipped CSV files, 6.59 GiB) using \emph{spark-shell} with default parameters. Dimensions are read from a 6.7 MiB CSV file. For the industry formats, dimensions are appended to data points from an in-memory cache. We also measure \sysg{}'s scalability on all six workers using bulk loading (B) and \emph{online analytics} (O) with aggregate queries executed on random time series using Segment View during ingestion. \sysg{} uses a specialized ingestor on one node, and Spark Streaming with 5 s micro-batches and one receiver per worker when distributed. We measure \sysg{} ingestion time while ingesting an unbounded time series on one node for 1.5 days, to show it is stable.

The results are shown in Figure~\ref{fig:query:ingestion}. \sysgonb{} and \sysgona{} create the same 1761 groups with 1.99 time series on average. Due to its better compression, \sysgona{} ingests 2.16--\emph{13.7}x faster than the other formats. \sysgoff{} is already 1.89x faster than \sys{} due to our optimizations (see Table~\ref{table:enhancements}). On one node, \sysg{} performed equivalently using spark-shell's default configuration and the cluster's configuration. On six nodes, \sysg{} achieves a 4.98--5.56x speedup for bulk loading and a 4.68--5.06x speedup for online analytics. The ingestion rate is stable and increases by 3\% over 1.5 days; short-term ingestion rate drops match JVM garbage collections.

\textbf{Effect of Error Bound and Grouping \label{sec:evaluation:experiments:eg}}
Compression is evaluated using EP, EF, and HD. $\epsilon$ is  0\%, 1\%, 5\%, and 10\% for \sys{} and \sysg{}, and 0\% for the industry formats. For EP, \texttt{Measure} determines correlation better than \texttt{Production}, whose weight thus is decreased to only group time series with equal \texttt{Production} members by \sysgona{}. For each data set we show the storage and model types used, and actual average error as $(\sum_{n=1}^{|DP|} |rv_n - av_n| / \sum_{n=1}^{|DP|} |rv_n|) $ $\times 100$ where $DP$ is ingested data points, $av_n$ is the $n$th approximated value, and $rv_n$ is the $n$th real value. As EP, EF, and HD contain correlated time series, grouping should reduce the storage used. We compare our grouping methods with a baseline value-based method that groups time series with equivalent min and max values (computed offline). Last, using instrumentation, we find the overhead of both static grouping, and dynamic splitting and merging of groups during ingestion, see Section~\ref{sec:partitioning}.

EP results are seen in Figure~\ref{fig:error:ep} with \sysgonb{} correlation set as \texttt{Production 0, Measure 1 ProductionMWh} as EP has multiple energy production measures per entity. \sysgonb{} creates 39,164 groups with an average size of 1.16 from 45,353 time series in only 17.86 minutes, while \sysgona{} creates 35,643 groups with an average size of 1.27 in only 21.13 minutes. Recall that \emph{static grouping is only run once before ingestion begins}. Splitting and merging use at most 1.19\% of the run-time. \sysg{} uses up to 16.2x less storage than the industry formats, while the highest average error is only 0.34\% ($\epsilon=10\%$). \sysgonb{} uses 1.44--1.56x less storage than \sysgoff{}, while \sysgona{} uses 1.18--1.37x less storage than \sysgoff{} as all measurements in the same category on each entity are grouped.

EF results are seen in Figure~\ref{fig:error:eh} with \sysgonb{} correlation set as \texttt{0.4166667} which groups the same measurement. \sysgonb{} creates 26 groups of average size 7.58 from 197 time series in 0.02 s; \sysgona{} creates 50 groups of average size 3.94 in 0.03 s. Splitting and merging use at most 1.88\% of the run-time. \sysg{} uses up to \emph{113}x less storage than the other formats, while the highest average error is only 1.72\% ($\epsilon=10\%$). \sysgonb{} uses 1.60--1.96x less storage than \sysgoff{}, while \sysgona{} uses 1.37--1.75x less storage than \sysgoff{} as the same measurements in a park are grouped.

For HD \sysgonb{} is not included in Figure~\ref{fig:error:hd} as \sysgona{} outperformed our manual attempts. \sysgona{} creates 132 groups of average size 2.5 from 330 time series in 0.05 s. Splitting and merging use at most  0.014\% of the run-time. \sysg{} uses up to 48.08x less storage than the other formats, while the highest average error is only 0.15\% ($\epsilon=10\%$). \sysgoff{} uses 1.36--2.54x less storage than \sysgona{} as prices for each pair are grouped.

Surprisingly, \sysgoff{} performs like \sys{} despite using the more restrictive PMC-Mean and not PMC-MR (see Table~\ref{fig:multi:ingestion}).

Figure~\ref{fig:error:models:ep}--\ref{fig:error:models:hd} show that all model types are used for all data sets, and that \sysg{} adapts to groups by using \ac{fb} more. This is expected as \emph{all} time series in an entire group must exhibit a constant or linear pattern to efficiently use \ac{pmc} or \ac{swing} with \ac{mmgc}, in contrast to \ac{mmc}.

When grouping EP with the baseline value-based method, 8,487 groups of average size 5.34 are created.
As \sysg{} stores gaps in 64 bits, groups larger than 64 are split. The storage required is 8.00--6.49 GiB ($\epsilon=0$--$10\%$). As EP contains no \textit{Location} dimension, our methods cannot group time series from co-located entities which might explain why the baseline is slightly better. For EF, 71 groups of average size 2.77 are created, using 1.09--0.89 GiB ($\epsilon=0$--$10\%$), and thus the baseline is slightly worse than our methods despite both grouping time series with similar measures. Last, for HD only 252 groups of average size 1.31 are created and a few groups had to be split. The storage required is 2.66--0.52 GiB ($\epsilon=0$--$10\%$), so the baseline again performs worse than our methods. In general, the baseline (value-based grouping) is worse than our grouping methods when considering all data sets, and furthermore requires the full data set to be analyzed offline.

\textbf{Automatic and Distance-based Grouping}
We evaluate distance-based grouping by ingesting all data sets with distances up to $0.50$. The number of dimensions and levels limit the possible distances, e.g., EP's are in $0.25$ increments.
The results for EP and EF are shown in Figure~\ref{fig:error:dist}. For HD only \texttt{0.3333334} is below $0.5$, and using \texttt{0.66666667} uses 1.39--6.33x more storage than \sysgoff{}. For distances above zero 35643--6757 groups of average size  1.27--6.71 are created for EP, and 65--26 groups of average size  3.03--7.58  for EH. As expected, only the lowest distance decreases  storage for all data sets compared to \sysgoff{}. Thus, our automatic grouping method is sound and users can exploit distance-based grouping as using \texttt{0.4166667} uses less storage than \texttt{auto} for EF.

\textbf{Scale-out}
Using L-AGG we evaluate the scalability of all formats on the cluster, and of \sysgonb{} on 1--32 Standard\_D8\_v3 Azure nodes (chosen based on Spark, Cassandra, and Azure documentation). Both use the same configuration, but Spark can use 50\% of each node's memory on Azure as no crashes occur. EP is duplicated until the data ingested by each node exceeds its memory, so \sysgonb{} cannot cache it all in memory. The values of each duplicated data set are multiplied with a random value in the range [0.001, 1.001) to avoid repeated values. Queries are run using an appropriate method: InfluxDB’s Java Client (J), ModelarDB’s Segment View (S) and Data Point View (DP), and a Spark Data Frame (F) for Cassandra, Parquet, and ORC.

Cluster results are shown in Figure~\ref{fig:query:lagg}. \sysgonb{} outperforms almost all other formats:  Parquet is only 1.66x faster despite its column-oriented layout which favors S-AGG. However, Parquet is 2.41x slower than \sysgonb{} to ingest data, does not support online analytics, and uses up to 11.6x more storage for EP. InfluxDB runs out of memory despite only sharing with the OS. In~\cite{db:2018:modelardb} we show that \sys{} outperforms InfluxDB at scale. The average query result error is only 0.024\% for \sysgoff{}, \sysgona{}, and \sysgonb{}. Azure results are shown in Figure~\ref{fig:query:scale}. \sysgonb{} scales linearly for S and DP, which is expected as \sysg{} assigns each time series group to a node so shuffling is avoided.

\textbf{Additional Query Processing Performance}
To further evaluate \sysg{} query performance we run S-AGG, P/R, and M-AGG on EP and EH on the cluster. \sys{} cannot run M-AGG as dimensions are not supported, nor can InfluxDB as dynamically sized time intervals are not supported~\cite{evaluation:2018:influxdb:durations}.

S-AGG results are shown in Figures~\ref{fig:query:sagg:ep}--\ref{fig:query:sagg:eh}. For EP, \sysg{} is much faster than Cassandra and only slightly slower than the rest. \sysgoff{} is slightly slower than \sys{} so supporting queries on groups adds a very small overhead on EP's short time series. \sysgoff{} and \sysgonb{} perform similarly despite \sysgonb{} reading a group for 25.7\% of the \texttt{Tid}s. The average query result error is only 0.043\% for \sysg{}. For EF, \sysgoff{} is only significantly slower than Parquet (4.38x) due to its column-oriented layout, but Parquet ingests 2.12x slower and uses 41.03--43.82x more storage for EF. Also, despite using significantly less storage, the average query result error is only 0.033\% for \sysgoff{}, 0.26\% for \sysgona{}, and 0.28\% for \sysgonb{}.
EF has long time series. As the average group size grows, queries must read larger groups to get a single time series and thus slow down. Users can thus prioritize query performance (for monitoring) or compression (for archiving).

P/R is not \sysg{}'s intended use case but included for completeness. For EP, \sysgonb{} is faster than Parquet (4.55x) and \sys{} (1.14x), but slower than InfluxDB (20.18x), Cassandra (3x), and ORC (1.52x). For EF, \sysgoff{} is faster than Cassandra (1.34x) and Parquet (1.05x), but slower than InfluxDB (766x), ORC (28.35x), and \sys{} (1.47x). While InfluxDB is faster for P/R, it cannot execute L-AGG and M-AGG, is 5.13x slower to ingest data, and uses up to 6.49x more storage. The average query result error is 0.012--2.01\% for \sysg{}. Thus, \sysg{} is competitive with the big-data formats. EF grouping has a trade-off between storage and performance.

M-AGG results are shown in Figures~\ref{fig:query:magg:ep:one}--\ref{fig:query:magg:eh:two}. For EP, M-AGG-1 queries \texttt{GROUP BY} \emph{category} which matches the groups created by \sysgonb{}. Thus, \sysgonb{} only reads the time series required for each query, making it 1.52--45.14x faster than the other formats. M-AGG-2 queries \texttt{GROUP BY} \emph{concrete} and as \sysgonb{} can execute queries on each time series in a group, unlike simple aggregates, it is 1.97--49.52x faster. The average query result error is 0.0027\% for \sysg{}. For EH, M-AGG-1 queries \texttt{GROUP BY} \emph{park} and \sysgoff{} is 3.17--\emph{630}x faster, while it is 2.93--579x faster for M-AGG-2 where the queries \texttt{GROUP BY} \emph{entity}. The average query result error is 0.014--0.17\% for \sysg{}. For EF, \sysgonb{} is a trade-off as it groups all energy production measures together.

\textbf{Summary of Evaluation}
Compared to the other formats, \sysg{} provides a high and stable ingestion rate (up to \emph{13.7}x faster) due to an efficient ingestion method and high compression (up to \emph{113}x less storage required) while supporting online analytics. The high compression is due to \algo{} compressing correlated time series together in groups using multiple model types. These groups are created automatically by \sysg{} (\sysgona{}) or by users (\sysgonb{}). \algo{} uses suitable model types for each data set and error bound pair. \sysg{} is up to 50.94x, 497x, and \emph{630}x faster for L-AGG, S-AGG, and M-AGG, respectively, due to model-based query processing, while still competitive for P/R. We have also shown that \sysg{} scales linearly.
 \section{Related Work}\label{sec:rw}
Surveys exist about models for time series~\cite{rw:2013:model:survey, rw:2013:model:evaluation}, compression of sensor data~\cite{rw:2013:wsn, rw:2018:compression:survey}, signal processing in the energy domain~\cite{rw:2018:dsp}, Hadoop-based OLAP~\cite{rw:2017:olap:survey} and \acp{tsms}~\cite{rw:2017:tsms:survey}.

\textbf{Group Compression:}
Compression of correlated time series is mainly used for sensor data acquisition~\cite{rw:2013:wsn}. Yang et al.~\cite{rw:2014:cloud}, e.g., cluster time series based on regression models and suppress transfer of similar data for each cluster.
Methods for \ac{tsms}s have been proposed.
Gamps~\cite{rw:2009:gamps} approximates time series using constant functions and then relaxes the error bounds before compressing them together.
MTSC~\cite{rw:2018:mtsc} uses graph-based methods to partition time series into correlated groups, uses constant functions for compression, and represents each group as a base signal and offsets.
CORAD~\cite{rw:2019:corad} is a lossy dictionary compression algorithm. Each segment is encoded as values from a pre-trained dictionary and deduplication is used for correlated segments.
Some methods use groups for other purposes than improving compression. Sprintz~\cite{rw:2018:sprintz} uses groups so decompression can use SIMD, while EdgeDB~\cite{rw:2019:edge} uses groups to reduce the number of I/O operations.
Other domains also use group compression, e.g., Elgohary et al.~\cite{correlation:2018:boehm:journal} compress correlated features for machine learning together.

\textbf{Multi-Model Compression:}
\ac{mmc} was proposed in~\cite{rw:multi:2011:towards}. Model types fit models to data points in parallel until they all fail, the model with the best compression is stored.
The Adaptive Approximation (AA) algorithm~\cite{rw:multi:2015:indexable} fits models in parallel and creates segments as each model type fails. After all model types fail, segments from the model type with the best compression are stored.
In~\cite{rw:multi:2015:grid} regression models are fitted in sequence with coefficients incrementally added. The model providing the best compression is stored when a user-defined number of coefficients is reached.
AMMMO~\cite{rw:multi:2020:mi} only uses lossless compression methods. Each segment is analyzed to create a set of candidate methods, and the best method for each data point is then selected from this set.

\textbf{Model-Based Data Management Systems:}
DBMSs with support for models have been proposed.
MauveDB~\cite{rw:mbdm:2006:mauvedb} integrates models into an RDBMS using views, to support data cleaning without exporting data to another application.
FunctionDB~\cite{rw:mbdm:2008:functiondb} natively supports polynomial function models and evaluates queries on these if possible.
Plato~\cite{rw:mbdm:2015:plato} uses models for data cleaning and allow user-defined model types to be added. Recent work~\cite{rw:mbdm:2020:plato} on Plato focuses on providing tight deterministic error guarantees for a set of queries on models.
Guo et al.~\cite{rw:mbdm:2014:kv-index} combine an in-memory tree index, a distributed key-value store, and MapReduce to store and query models in a distributed system.
\sys{}~\cite{db:2018:modelardb} is a model-based \ac{tsms}, with a \ac{mmc} compression method that supports user-defined model types, that integrates with Spark and Cassandra.

\textbf{\ac{mmgc}, \algo{}, and \sysg{}:}
In contrast to existing model-based compression algorithms~\cite{rw:2009:gamps,rw:2018:mtsc,rw:2019:corad,rw:multi:2011:towards,rw:multi:2015:indexable,rw:multi:2015:grid,rw:multi:2020:mi} and systems~\cite{rw:mbdm:2006:mauvedb,rw:mbdm:2008:functiondb,rw:mbdm:2015:plato,rw:mbdm:2014:kv-index,db:2018:modelardb,rw:mbdm:2020:plato}, \sysg{} uses multiple model types to compress time series groups. This unifies \ac{mmc} and \ac{mgc}, and creates the new \ac{mmgc} category of model-based compression methods with \algo{} being the first. In addition, \sysg{} can automatically group correlated time series using only their dimensions, and provides primitives for user to effectively specify groups. As groups are created before ingestion, workers receive data directly from the time series.
Compared to \algo{}, the compression methods used for smart meter data~\cite{rw:2018:compression:survey} (e.g., Huffman Encoding and LZ-based methods) are not optimized for regular time series with gaps. They also only use one representation each, making them unable to adapt as time series change~\cite{rw:2013:model:survey}, while \algo{} compensates for the weaknesses of one model type by using another at run-time. In terms of compression, Figure~\ref{fig:error:ep}--\ref{fig:error:hd} clearly shows that \algo{} outperforms LZ4 as used by Cassandra. Also, \sysg{}'s extension API allows users to add the methods as new model types without recompiling the system.
In summary, \sysg{} provides state-of-the-art compression and query performance by storing groups of correlated time series as compressed models and executing OLAP queries on these.
 \section{Conclusion \& Future Work}\label{sec:conclusion}
Motivated by the need to efficiently manage huge data sets of time series from reliable sensors, we presented several novel contributions: (i) The \emph{\acf{mmgc}} category of model-based compression methods for time series. (ii) The first \ac{mmgc} method \emph{\algo{}} and model types extended to compress time series groups. (iii) Primitives so users effectively can group time series, and based on these, an automated method using the time series dimensions. (iv) Algorithms for executing simple and multi-dimensional aggregate queries on models of time series groups. We added our methods to the model-based \ac{tsms} ModelarDB. We named this version \emph{ModelarDB\textsubscript{+}}. Our evaluation showed that compared to widely used systems, ModelarDB\textsubscript{+} provides up to: 13.7x faster ingestion due to high compression, 113x better compression due to \algo{}'s adaptivity, and 630x faster aggregates as they are computed from models.

In future work, we will simplify using ModelarDB\textsubscript{+} and increase its query performance by: (i) Developing indexing techniques exploiting that data is stored as models. (ii) Developing query and cluster aware grouping and partitioning methods. (iii) Supporting high-level analytical queries, e.g., similarity search, on models. (iv) Removing or inferring parameter arguments.

\section{Acknowledgments}
This research was supported by the DiCyPS center funded by Innovation Fund Denmark, the GOFLEX project EU grant No 731232, and Microsoft Azure for Research. We also thank our data providers for the real-life data sets and domain insights.

\bibliographystyle{IEEEtran}

\end{document}